\documentclass[aps,prl,twocolumn,showpacs]{revtex4}
\usepackage{epsfig}
\usepackage{graphicx}
\usepackage{amsmath}

\begin{document}

\DeclareGraphicsExtensions{.eps,.EPS}

\title{Cooling of a Bose-Einstein Condensate by spin distillation}
\author{B. Naylor$^{1,2}$, E. Mar\'echal$^{1,2}$, J. Huckans$^{1,3}$, O. Gorceix$^{1,2}$,  P. Pedri$^{1,2}$, L. Vernac$^{1,2}$, B. Laburthe-Tolra$^{1,2}$}

\affiliation{$^{1}$\,Universit\'e Paris 13, Sorbonne Paris Cit\'e, Laboratoire de Physique des Lasers, F-93430, Villetaneuse, France\\
$^{2}$\,CNRS, UMR 7538, LPL, F-93430, Villetaneuse, France\\
$^{3}$\, Department of Physics and Engineering Technology, Bloomsburg University of Pennsylvania, Bloomsburg, PA 17815, USA}

\begin{abstract}

We propose and experimentally demonstrate a new cooling mechanism leading to purification of a spinor Bose-Einstein Condensate (BEC). Our scheme starts with a BEC polarized in the lowest energy spin state. Spin excited states are thermally populated by lowering the single particle energy gap set by the magnetic field. Then these spin-excited thermal components are filtered out, which leads to an increase of the BEC fraction. We experimentally demonstrate such cooling for a spin 3 $^{52}$Cr dipolar BEC. Our scheme should be applicable to Na or Rb, with perspective to reach temperatures below 1 nK.
\end{abstract}
\date{\today}
\maketitle

In the last three decades, laser cooling and evaporative cooling have led to major advances in atomic and molecular physics, in particular in the fields of precision measurements, atomic clocks, and quantum degenerate gases \cite{Perrin2011,Chu2002}. Nowadays, the hope to study magnetic correlations of atoms in optical lattices \cite{Greif2013,HuletNature}, and the possible connections to exotic superconductivity are major motivations to obtain systems with lower entropies than currently available \cite{BookQS}. It is therefore important to find new ways to remove entropy in degenerate quantum Bose or Fermi gases loaded in optical lattices \cite{ProposalEntropy,DSK09,Medley2011,QDistillation}.

In typical lattice experiments, atoms are first cooled by evaporative cooling, and then loaded in the periodic potential. Unfortunately, evaporative cooling ceases to be efficient when the temperature is significantly smaller than the interaction energy, which currently sets an ultimate limit for entropy \cite{DeMarcoRPP2011}. A number of alternative cooling schemes have been suggested and studied \cite{RevBECFormation,KetterleCooling, BECconstantT, ZeemanCooling, demagnetizationCooling, Stellmer2013} but up to now the best phase space densities are still due to evaporative cooling. Here, we propose to use the spin degrees of freedom to efficiently store and remove entropy in partially Bose-condensed gases to reach temperatures below the current limitations set by evaporative cooling.

\begin{figure}
\centering
\includegraphics[width= 3 in]{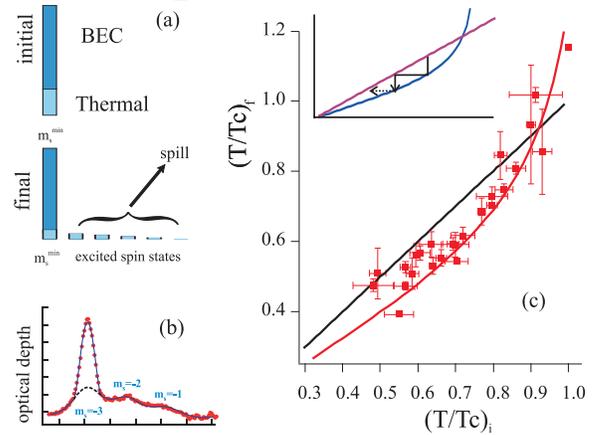}
\caption{\setlength{\baselineskip}{6pt} {\protect\scriptsize a) Principle of the proposal. b) Density profiles after depolarization and Stern Gerlach procedure showing a $^{52}$Cr BEC forming only in $m_s=-3$. Circles are experimental data, the solid line is result of bimodal fit, and the dashed line only fits thermal fractions.  c) Purification of a $^{52}$Cr BEC. After depolarization occurring at a B field of 1 mG, we measure the final condensate fraction and hence the effective final reduced temperature (red squares), as a function of the initial reduced temperature. Error bars show statistical uncertainties. The (red) solid line is the result of our model. The (black) straight solid line (of unity slope, crossing the origin) is a reference (i.e. no cooling). The inset illustrates cooling by successive cycles.}}
\label{principle}
\end{figure}

Our proposal starts with a sample polarized in the lowest energy spin state. We engineer a thermodynamic cycle (see fig. \ref{principle}a)) wherein the magnetic field is first lowered to trigger depolarization of the gas, and the resultant spin-excited states are then filtered out of the trap. As shown in fig. \ref{principle}c), this cycle reproduces a BEC polarized in the lowest energy state with an increased condensate fraction, provided the initial thermal fraction is low enough. The gain in phase space density directly follows from Bose statistics: as the condensate forms in the lowest energy single-particle spin state \cite{ThermoSpinor}, spin filtering of the excited spin states obtained after depolarization introduces a loss which is specific to thermal atoms. At low magnetic field, spin filtering typically leads to a decrease of the thermal fraction, and hence of entropy, by a factor $\simeq2$. The cycle can then be repeated. We find that this cooling efficiency has no fundamental limit other than that associated with technical noise.

To demonstrate this new cooling strategy, we consider two specific systems. The first system is a chromium dipolar gas, which is sensitive to the linear Zeeman effect, because magnetization is free \cite{ThermoSpin3}. Bose-Einstein condensation occurs in the lowest energy spin state $m_s=-3$ (above a critical B field, see below). Then, thermal population of $m_s>-3$ spin-excited states due to dipole-dipole interactions (see fig. \ref{principle}b)) can be used for cooling. The cooling limit arises from technical difficulties at controlling the magnetic field at the 100 $\mu$G level. Despite these difficulties, this paper provides a proof-of-principle experiment, demonstrating the efficiency of spin filtering to purify a chromium BEC in an optical dipole trap.

The second system is a spinor $s=1$ BEC such as Rb or Na, prepared in $m_s=0$. Spin-changing collisions associated with the difference in scattering lengths in the molecular potentials $S=0$ and $S=2$ redistribute population between $m_s=0$ and $m_s=\pm 1$ at constant magnetization.
The (positive) quadratic Zeeman effect provides an energy shift $q$ between a pair of atoms in the $m_s=0$ state and a pair of atoms in states $m_s=\pm 1$, which favors BEC in $m_s=0$. Thermal population of spin-excited states is possible as long as $k_B T > q$, which sets a practical limit to cooling in the pK range. This limit ensures that spin fluctuations of mesoscopic polar BECs are negligible \cite{Sarlo2013}.

Before discussing our experimental results, we now turn to our theoretical model. The suggested experiments start with a finite temperature polarized BEC in the lowest energy spin state $m_s^{min}$. The magnetic field is then rapidly (diabatically) reduced to allow depolarization. To compute the equilibrium state after depolarization, we assume thermodynamic equilibrium (insured by collisions), and conservation of the total atom number $N_{tot}$ and total energy. We use the thermodynamics of non-interacting atoms, assuming that mean-field interactions can be neglected when evaluating the equilibrium state, so that spin-dependent Bose occupation factors read:

\begin{equation}
f_{k,m_s}(\mu)=  \frac{1}{\exp{\left[\beta\left(\epsilon_{k,m_s}-\mu \right)\right]}-1}
\end{equation}
where $\mu$ is the chemical potential, $\beta=1/k_BT$, and $\epsilon_{k,m_s}$ is the single-particle energy of the (trapped) states labeled by the index $k$, in the Zeeman state $m_s$. When magnetization is free (case of dipolar particles) $\epsilon_{k,m_s}$ includes the linear Zeeman effect, whereas when magnetization is fixed (as for example for Rb and Na atoms), the linear Zeeman effect is gauged out and $\epsilon_{k,m_s}$ only includes the quadratic Zeeman effect $q$. In both cases, we specify that the lowest energy spin state $m_s^{min}$ ($m_s^{min}=-3$ for Cr with free magnetization, $m_s^{min}=0$ for Rb or Na at zero magnetization) has $\epsilon_{k\rightarrow0,m_s^{min}} =0 $ for the lowest trapped states $k\rightarrow0$ (which sets the otherwise arbitrary energy scale). We assume that a BEC is present in state  $m_s^{min}$, so that $\mu=0$.

\begin{figure}
\centering
\includegraphics[width= 3.5 in]{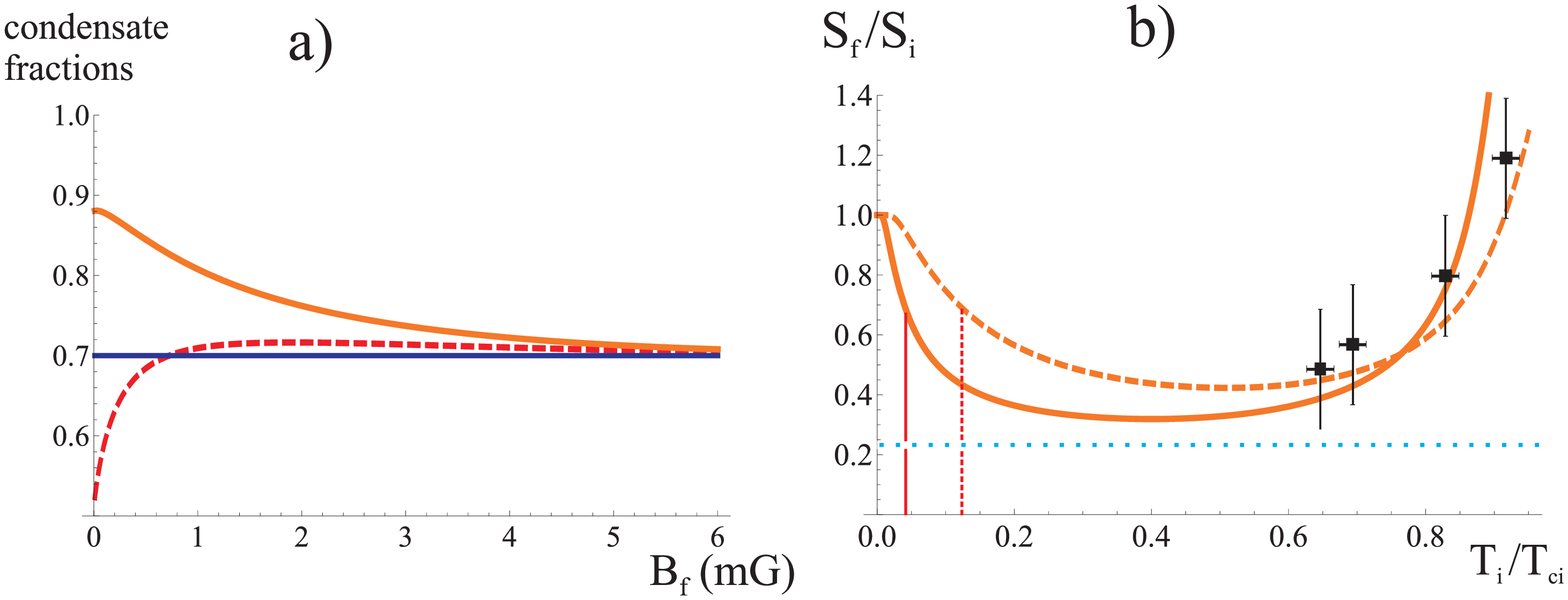}
\caption{\setlength{\baselineskip}{8pt} {\protect\scriptsize (Color online) Theoretical results for the s=3 $^{52}$Cr BEC. a) Condensate fractions for $f_{i}=0.7$ after depolarization, with (solid, orange) or without (dashed, red) spilling, as a function of B$_f$. b) Ratio of final (after spilling) and initial mean entropy per atom as a function of initial temperature, for B$_f$=0.3 mG (dashed) and B$_f$=0.1 mG (solid). The horizontal dotted line corresponds to the maximal gain, reached at B$_f$=0 and $T_i=0$ (given by eq. (\ref{epsilons0})). The vertical lines indicate the qualitative limit for cooling, set by $g \mu_B$B$_f\simeq k_B T$. The dots (black) correspond to experimental data taken at B$_f=0.2\pm0.1$ mG, including statistical uncertainties.}}
\label{FigTheory}
\end{figure}

In the case of a 3D harmonic trap, and in the thermodynamic limit, the thermal population in each spin state reads:
\begin{equation}
N_{m_s}= \sum_{k} f_{k,m_s}(0) \approx g_3\left(e^{-\beta \epsilon_{0,m_s}} \right)  \left(\frac{k_B T}{\hbar \omega}\right)^3
\end{equation}
with $g_n$ the polylogarithm function of order $n$, and $\omega$ the geometric average of the angular trapping frequencies of the 3D harmonic trap. Additional atom are condensed in state $m_s^{min}$.

The total energy of the system $\sum_{k,m_s} f_{k,m_s}(0)\epsilon_{k,m_s}$ is the sum of trap (kinetic and potential) and magnetic energies:

\begin{eqnarray}
E_{trap} &  \approx & \sum_{m_s} 3 k_B T g_4\left(e ^{-\beta \epsilon_{0,m_s}}\right) \left(\frac{k_B T}{\hbar \omega}\right)^3 \nonumber\\
E_{mag} & = & \sum_{m_s}N_{m_s}\epsilon_{0,m_s}
\end{eqnarray}

The final temperature $T_f$ and BEC atom number $N_{c,f}$ are thus derived, and can be compared to initial values $T_i$ and $N_{c,i}$. Any atom in spin-excited state $m_s \neq m_s^{min}$ can be removed (by means of magnetic field gradients, micro-wave transitions, and/or a resonant push laser beam). We define two post-depolarization BEC fractions, with or without spin filtering the excited states, which are respectively  $f_{2}=N_{c,f}/(N_{c,f}+N_{m_s^{min},f})$, and $f_{1}=N_{c,f}/N_{tot}$ ($N_{tot}=N_{c,i}+N_{m_s^{min},i}$).

As shown in fig. \ref{principle}c), we find that this procedure should lead to an increase in condensate fraction $f_{2}$ only when the initial condensate fraction $f_{i}=1-\left(T_i/T_{ci}\right)^3$ is large enough. We interpret this complex behaviour as the consequence of the competition between two effects. (i) As population in spin-excited states is purely thermal, spin filtering leads to purification of the BEC. (ii) As the gas depolarizes and the number of thermal atoms in $m_s^{min}$ decreases, the BEC must melt to maintain saturation.  This competition can lead to an increase in condensate fraction, because BEC atoms have zero energy. Hence, melting of the BEC cools the thermal gas in $m_s^{min}$, which can then be saturated at a lower temperature, as already observed in \cite{Cornell2003} and \cite{QJT}.

At low enough $T$, the cooling process becomes most efficient for B$\rightarrow 0$ ($q\rightarrow 0$), and in that limit we obtain the following simple expression:

\begin{equation}
1-f_2=\frac{1-f_i}{(2s+1)^{3/4}-2s(1-f_i)}
\label{epsilons0}
\end{equation}
Therefore, at zero B field, while $f_1=1-(2s+1)^{1/4}(1-f_i)<f_i$, $f_2>f_i$ for large enough $f_i$ (if $f_i > 0.45$ for $s=3$).

It is interesting to assess the efficiency of this cooling procedure. The best way to do so is to consider the reduction of entropy after each cooling cycle. Below $T_c$, all entropy is contained in the thermal fraction, and each thermal atom has a temperature-independent entropy of 3.6 $k_B$ \cite{Huang}. As a consequence, selectively removing thermal atoms is an excellent way to remove entropy. For B=0, $\frac{S_f}{S_i}=\frac{1-f_2}{1-f_i}\rightarrow \frac{1}{(2s+1)^{3/4}}$ when $f_i\rightarrow 1$, with $S_{i,f}$ the mean entropy per atom before depolarization, and after spilling, respectively. In fig \ref{FigTheory}b) the entropic efficiency of one cooling cycle is shown for different values of the B field, as a function of $T_i$, in the case of $s=3$ Cr. $\frac{S_f}{S_i}$ remains significantly smaller than 1 even at very low $T_i$, provided a low enough B field: a reduction of entropy by a factor of at least two is obtained until $k_B T_i\simeq g \mu_B$B. This corresponds to $T_i\simeq0.1 \times T_{ci}=30$ nK in our experiment. With $s=1$ Na or Rb, this limit can be pushed to even lower $T_i$. For example for Na, with an energy difference of 280 Hz/G$^2$ between $ \left \{ m_s=0;m_s=0\right \}$ and $\left \{m_s=-1;m_s=1 \right \}$, temperatures in the 100 pK range can be achieved with B fields below $\simeq100$ mG.

\begin{figure}
\centering
\includegraphics[width= 3.0 in]{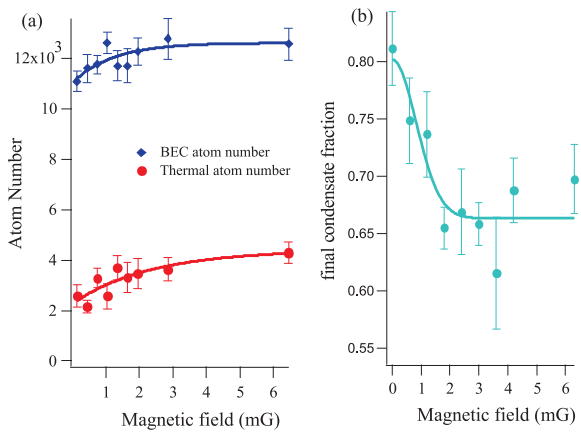}
\caption{\setlength{\baselineskip}{8pt} {\protect\scriptsize (Color online) Experimental results for spin spilling with the $s=3$ $^{52}$Cr BEC, for an initial BEC fraction of 0.65, and different values of B$_f$.  a) Number of atoms after spin spilling: in the BEC (top), in the thermal component.  b) BEC fraction deduced from a). The solid lines are guides for the eye. }}
\label{Manip1}
\end{figure}

We have also estimated the effect of interactions, by including in $\epsilon_{k,m_s}$ the effect of spin-dependent contact interactions within the Bogoliubov approximation \cite{Uedareport}. We find that interactions between particles do not modify the general picture described above, as long as the BEC remains polarized in a well defined Zeeman state. For the case of Na, spin-dependent interactions favor a polar $m_s=0$ BEC. For Rb, at typical densities of 10$^{14}$ at.cm$^{-3}$, a quadratic energy shift of typically 10 Hz, provided by a magnetic field of 380 mG, is sufficient to insure that the BEC is polar. For the case of chromium (Rb), the BEC is polarized in $m_s=-3$ ($m_s=0$) above a critical magnetic field B$_c$ set by spin-dependent interactions \cite{PhaseCr,Depolarization}. In the experiment reported here, B$_c\simeq 100$ $\mu$G.

Finally, we also find a subtle effect associated with demagnetization cooling (a process previously demonstrated in the group of T. Pfau in the thermal regime \cite{demagnetizationCooling}). When the experiment is performed at non-zero magnetic field, the cooling associated with the exchange between magnetic and kinetic energy can lead to an increase in the BEC atom number \textit{even without spin filtering}: $N_{c,f}>N_{c,i}$, and hence $f_1>f_i$, as shown in fig \ref{FigTheory}a). However, this effect is small compared to the two major effects described above (and smaller than our experimental error bars). The increase in BEC atom number occurs despite the necessary increase in total entropy, because spin-excited states have more entropy per particle than the saturated (thermal) lowest energy state. This bears similarity with the reversible production of a BEC when tuning the anharmonicity of a trap \cite{RevBECFormation}.

We now present our experimental results demonstrating purification of a $^{52}$Cr BEC. Condensation is obtained in the absolute ground state $m_s=-3$, by evaporative cooling in a crossed optical dipole trap obtained from a 100 W, 1075 nm, IR laser \cite{double trap}. The final trap frequencies at the end of evaporation are varied in order to change the initial BEC fraction $f_i$ at the beginning of the spilling process. For example, for $\omega_{x,y,z}=2\pi\times(250,300,215)$ Hz (measured through parametric excitation, with $5\%$ uncertainty), we find $f_i=0.5$. This is in agreement with a critical temperature for BEC $T_c=310$ nK for $2.10^4$ atoms, and a measured temperature of 250 nK.

The magnetic field, which is initially B$_i$=40 mG, is then lowered in 50 ms to a final value B$_f$, low enough to trigger depolarization ($g\mu_B$B$_f \simeq k_BT$). To calibrate B$_f$, we first search for the B field where maximal depolarization occurs, by varying all three B components with steps of 100 $\mu$G, which defines the zero B field. We then leave two components of the B field unchanged, and vary one component, which is calibrated by use of radio-frequency spectroscopy. We then let the cloud evolve at B$_f$ for 150 ms. This is long enough for inelastic dipolar collisions to ensure depolarization, as the typical rate for these collisions is 15 s$^{-1}$ for a magnetic field in the 0-1 mG range \cite{PRAreldip}.

\begin{figure}
\centering
\includegraphics[width= 2.5 in]{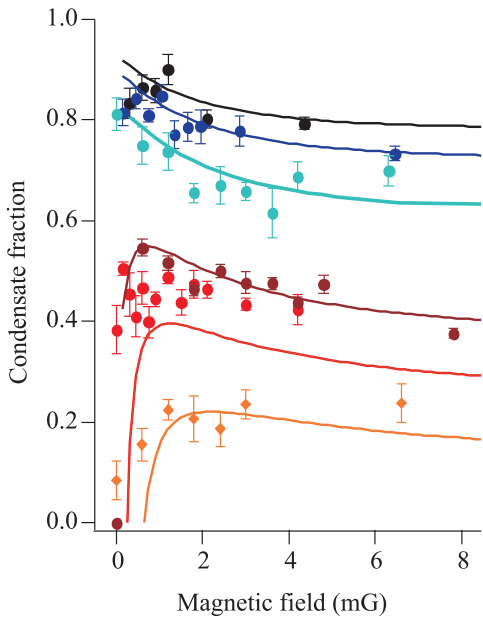}
\caption{\setlength{\baselineskip}{8pt} {\protect\scriptsize (Color online) Experimental results for spin spilling with the $s=3$ $^{52}$Cr BEC, for different initial BEC fractions $f_i$. The BEC fraction after spilling, $f_2$, is plotted as a function of B$_f$. Lines are predictions of our non-interacting model. For large B$_f$, $f_2$ reaches $f_i$.}}
\label{Manip2}
\end{figure}

The spin filtering procedure consists of i) raising the magnetic field to 30 mG; ii) changing the trap in order to spill out atoms with $m_s>-3$. For that we decrease the IR power and add a vertical B field gradient. In this trap, $m_s=-3$ atoms remain trapped, while $m_s>-3$ atoms fall, because optical and magnetic forces are insufficient to compensate gravity for these atoms. We checked through Stern-Gerlach analysis that indeed the losses induced by the new trap are negligible for $m_s=-3$, but almost total for $m_s>-3$. In addition, we checked that no evaporation is induced by the filtering procedure. The spin filtering is performed rapidly enough (in 60 ms) to ensure that no dipolar relaxation occurs. Finally, we recover the initial (purely) optical trap at B$_i$, and measure the new BEC fraction, $f_2$.

To accurately measure BEC fractions, we release the atoms from the trap and take an absorption picture after a time of flight of 5 ms. We perform a two-stage analysis of the images. A first fit is used to measure the total number of atoms, and deduce the critical temperature $T_c$. A second bi-modal fit is then performed, where the temperature is a free parameter which sets both the width of the gaussian describing thermal atoms, and the BEC fraction following non-interacting Bose thermodynamics predictions. This procedures ensures measurement of thermal fractions down to a few percent. Results are shown in fig.\ref{Manip1} for an initial temperature (corresponding to $f_i=0.65$) low enough that the spilling is efficient. At the lowest B$_f$, the spilling procedure induces a relative loss in thermal atoms larger than that for condensed atoms. This figure illustrates the competition between the two effects described above, which results here in an increase of the BEC fraction $f_2$.

Figure \ref{Manip2} shows our complete set of data, for different initial condensate fractions $f_i$. We observe different behaviors for $f_2$ at the lowest B$_f$. For the smallest $f_i$, $f_2$ gets smaller than $f_i$. On the other hand, when $f_i$ is large enough, $f_2$ gets significantly larger than $f_i$. All these results agree with our simple non-interacting model. In terms of entropy, the measured efficiency of the scheme (see fig \ref{FigTheory}) also agrees with our model; in particular, the reduction of entropy is more pronounced as $T_i$ decreases.

In conclusion, we propose an efficient cooling mechanism using on the spin degree of freedom. It is based on redistribution of entropy among the spin states, while recent theoretical proposals \cite{ProposalEntropy} or experiments rely on spatial redistribution \cite{DSK09}, as is for example the case for spin gradient demagnetization cooling \cite{Medley2011}, or distillation of singly-occupied sites in optical lattices \cite{QDistillation}. In addition, cooling by spin filtering can be repeated an arbitrary number of times. Since each cycle leads to typically a factor of two reduction in mean entropy per atom, we foresee that this scheme could indeed be a way to reach new regimes of deep degeneracy. As very pure condensates are obtained, measuring small thermal fractions will be a challenge. It will then be advantageous to use the spin degrees of freedom for thermometry, as explored in \cite{ThermoSpin3}, and discussed in \cite{SantosThermometry}: counting atoms in spin-excited states offers a background-free measurement, contrary to momentum distributions measurements. Finally, one of the interesting pending question is whether this scheme will help to remove entropy for BECs loaded in optical lattices, e.g. in the superfluid shells surrounding the Mott plateaux characteristic of the typical wedding cake distribution \cite{Svistunov2002}.

Acknowledgements: LPL is Unit\'e Mixte (UMR 7538) of CNRS and
of Universit\'e Paris 13, Sorbonne Paris Cit\'e.
We acknowledge financial support from Conseil R\'{e}%
gional d'Ile-de-France under DIM Nano-K / IFRAF, CNRS, and from Minist\`{e}re de
l'Enseignement Sup\'{e}rieur et de la Recherche within CPER Contract.

\end{document}